\documentclass[acmsmall]{acmart}
\usepackage{array}
\usepackage{geometry}
\AtBeginDocument{%
  \providecommand\BibTeX{{%
    \normalfont B\kern-0.5em{\scshape i\kern-0.25em b}\kern-0.8em\TeX}}}

\setcopyright{acmcopyright}
\acmJournal{PACMHCI}
\copyrightyear{2024}
\acmYear{2024}\acmDOI{XXXXXXX.XXXXXXX}

\begin{document}

%%
%% The "title" command has an optional parameter,
%% allowing the author to define a "short title" to be used in page headers.
%\title{Off the Cushion: Everyday Mindfulness and Technology Use}
\title[Beyond Meditation]{Beyond Meditation: Understanding Everyday Mindfulness Practices and Technology Use Among Experienced Practitioners}
%%
%% The "author" command and its associated commands are used to define
%% the authors and their affiliations.
%% Of note is the shared affiliation of the first two authors, and the
%% "authornote" and "authornotemark" commands
%% used to denote shared contribution to the research.

\author{Jingjin Li}
\affiliation{%
  \institution{Cornell University}
  \city{Ithaca}
  \state{New York}
  \country{USA}}
\email{jl3776@cornell.edu}

\author{Karen Anne Cochrane}
\affiliation{%
  \institution{University of Waterloo}
  \city{Waterloo}
  \country{Canada}}
\email{karen.cochrane@waterloo.ca}

\author{Gilly Leshed}
\affiliation{%
  \institution{Cornell University}
  \city{Ithaca}
  \state{New York}
  \country{USA}}
\email{gl87@cornell.edu}

%%
%% By default, the full list of authors will be used in the page
%% headers. Often, this list is too long, and will overlap
%% other information printed in the page headers. This command allows
%% the author to define a more concise list
%% of authors' names for this purpose.
\renewcommand{\shortauthors}{Li et al.}

%%
%% The abstract is a short summary of the work to be presented in the
%% article.
\begin{abstract}
  Mindfulness, a practice of bringing attention to the present non-judgmentally, has many mental and physical well-being benefits, especially when practiced consistently. Many technologies have been invented to support solo or group mindfulness practice such as mobile apps, live streams, virtual reality environments, and wearables. In this paper, we present findings from an interview study with 20 experienced mindfulness practitioners about their everyday mindfulness practices and technology use. Participants identify the benefits and challenges of developing long-term commitment to mindfulness practice. They employ various strategies, such as brief mindfulness exercises, social accountability, and guidance from teachers, to sustain their practice. While conflicted about technology, they adopt and appropriate a range of technologies in their practice for reminders, emotion tracking, connecting with others, and attending online sessions. They also carefully consider when to use technology, when and how to limit its use, and ways to incorporate technology as an object for mindfulness. Based on our findings, we discuss expanding the definition of mindfulness and the tension between supporting short- and long-term mindfulness practice. We also propose a set of design recommendations to support everyday mindfulness including such as through the lens of metaphor, reappropriating non-mindfulness technology, and bringing community support into personal practice.
\end{abstract}

%%
%% The code below is generated by the tool at http://dl.acm.org/ccs.cfm.
%% Please copy and paste the code instead of the example below.
%%


%%
%% Keywords. The author(s) should pick words that accurately describe
%% the work being presented. Separate the keywords with commas.

%%
%% This command processes the author and affiliation and title
%% information and builds the first part of the formatted document.
\maketitle

\section{Introduction}
Mindfulness, a practice of maintaining awareness by bringing attention to the present without judgment \cite{kabat2009full}, has been growing in popularity in the past decades. Scientific research has shown that mindfulness practice brings various mental and physical well-being benefits such as reduced stress and anxiety, improved emotion regulation, improved focus, and reduced chronic pain \cite{brown2003benefits}. Similar to other cognitive or physical exercises, obtaining these benefits relies on people's consistent practice. 

Mindfulness has been a growing trend in recent Human-Computer Interaction (HCI) research \cite{lukoff2020ancient, dauden2018evaluating, derthick2014understanding, dauden2020body, zhu2017designing, Tan2023-mindfulmoment, Li2022-beyond, Li2022-designing, Li2023-Co-design-mindfulness}, with a range of perspectives on what mindfulness is and motivations for the research \cite{terzimehic2019review}. Mindfulness HCI research also strives to design, develop, and evaluate digital technologies to support mindfulness practice. These include, for example, web-based mindfulness practice guidance \cite{krusche2013mindfulness}, wearables that detect stressful moments and deliver in-time mindfulness interventions \cite{kressbach2018breath}, and immersive virtual reality environments for mindfulness practice \cite{amores2016psychicvr, roo2017inner, prpa2018attending}. With a myriad of commercial technologies to support mindfulness (e.g., Calm, Headspace), HCI research also explores the features of mobile mindfulness meditation apps and conducted expert reviews of these apps \cite{lukoff2020ancient, dauden2018evaluating, schultchen2021stay}. Findings from these studies provide a comprehensive understanding of digital technologies, their affordances, benefits, and drawbacks. Research on people's daily experiences and technologies to support mindfulness practice has focused on novices in a mindfulness training program \cite{parsons2019designing}, or on experienced practitioners' formal meditative practice \cite{Markum20}. 

Despite the development of those mindfulness technologies and empirical studies examining their use, there is a gap in understanding how \textit{mindfulness practitioners}, who have been practicing mindfulness for a long time, incorporate or refrain from using these technologies in their daily practice. Understanding the practices of those long-term practitioners with sustained engagement with mindfulness has the impact of designing practical everyday mindfulness technologies that resonate with individuals' well-being needs in the long-term. 

To address this gap, we focus on understanding how mindfulness practitioners practice mindfulness both \textit{formally} and \textit{informally} in their daily activities. We examine their experienced benefits and challenges, and the technologies they use in their practice. We are especially interested in the potential of social computing technologies and how these technologies can support long-term mindfulness practice. In this study, we address the following research questions:
\begin{itemize}
    \item RQ1: What are the working definitions and characteristics of both informal and formal mindfulness practice from the practitioners' perspectives to give HCI researchers a richer understanding of holistic mindfulness practice?
    \item RQ2: What are the motivations, perceived benefits, and challenges faced by mindfulness practitioners in maintaining a regular practice? How can we learn from these insights to inform designs that support consistent mindfulness practice?
    
    \item RQ3: What roles do technologies play in practitioners' everyday mindfulness practice? How does it inform the design of future social computing technologies for mindfulness and mental well-being? 
\end{itemize}

We approach our research questions by conducting semi-structured interviews with 20 mindfulness practitioners who have been regularly practicing mindfulness for four years or more. Our findings point to a range of definitions for mindfulness and what it consists of, based on participants' everyday formal and informal mindfulness activities. Furthermore, short- and long-term physical and mental benefits motivated practitioners to practice it more consistently. Despite the benefits, all participants, even those with established long-term daily practice routines, felt that sticking to the practice was not easy. They adopted various strategies to maintain the practice, including practicing short mindfulness exercises, leveraging social accountability, and seeking help from a teacher. While being wary about relying on or including technology for mindfulness, experienced practitioners adopt and appropriate a range of technologies for their daily mindfulness practice. These included technologies that are not necessarily designed to support mindfulness practice, such as a calendar for reminders, spreadsheets for recording emotions, music and video streaming platforms, and video conferencing tools to connect with others and attend online meditation sessions. To support their practice, participants carefully considered their use of technology, when and how to limit its use, and ways to incorporate technology as an object for mindfulness.

Based on our findings, we propose that mindfulness technology should be designed in a way that supports mindfulness as a personally meaningful and ongoing process that incorporates practitioners' informal daily activities and technology habits. We discuss the tension between supporting short- and long-term mindfulness practice. Lastly, we propose a set of design recommendations to support everyday mindfulness including designing for mindfulness through the lens of metaphor, using a toolkit approach to design for mindfulness, reappropriating non-mindfulness technology for mindfulness, minimizing the negative effects of technology on mindfulness etc. We also highlight leveraging social computing tools to bring community support into personal mindfulness practice such as utilizing live stream mindfulness \cite{Li_Guo_LiveStream} for collective awareness. 

This work contributes to: (1) an in-depth and holistic understanding of everyday mindfulness practices from the perspectives of mindfulness practitioners, and their technology use and non-use in the practice. (2) a set of recommendations for the research and design of mindfulness technologies that are better situated and integrated into practitioners' daily lives, along  with opportunities for HCI researchers to design future technologies to support mindfulness practice.

\section{Background}
In this section, we review the current literature on mindfulness practice, the definitions, and the well-being benefits of mindfulness along with the current HCI and CSCW literature which looks at the relationship between mindfulness and technology. Through the literature review we highlight the current gap and articulate how our study fills it. 

\subsection{Definitions of Mindfulness}

The source of mindfulness practice is spiritual, and although mindfulness can be found in almost every religion, we look towards to Buddhist tradition to understand its origins. In Buddhism, mindfulness comes from the word \textit{Sati} which loosely means "to remember", "to recollect", or "to bear in mind" in the Pali language. It can be interpreted as being aware of one's sensations being illusions and not permanent \cite{bodhi2013does}. According to Buddhist faith, the only way to end or be free of suffering is by removing oneself from the craving and attachment to these sensations, whether they are pleasant or unpleasant. In this tradition, mindfulness is oriented to ethical and spiritual considerations of Buddhism as a religious faith and practice \cite{bodhi2013does, nilsson2016buddhist}. 

A more commonly used definition of mindfulness, \textit {paying attention to the present moment without judgment}, was popularized by Jon Kabat-Zinn, the pioneer of the eight-week Mindfulness-Based Stress Reduction (MBSR) Program. This therapeutic approach borrowed mindfulness from Buddhism, adapting it as a behavioral intervention to help people regulate their attention and emotions for mental and physical health-related outcomes \cite{kabat2009full}. In the past decades, mindfulness practice has been widely applied in clinical and non-clinical therapy groups in the West \cite{Creswell2017-sk}. 

Building on the therapeutic approach, psychology research has operationalized mindfulness as a two-component concept: (1) self-regulation of one's attention to the present moment, with (2) an orientation of curiosity, openness, and acceptance to this present experience \cite{Bishop2004-un}. Mindfulness can be viewed as a practice, a state, or a trait \cite{Kiken2015-la}. As a practice, mindfulness refers to a set of mental exercises to achieve the state of mindfulness, by bringing awareness to the present moment with an open and non-judgmental orientation. As a state, it refers to being in a mindful situation, where one is non-judgmentally aware of the present moment. And as a trait, it is one's capacity to easily get into a mindful state. Research has shown that increasing the frequency of state mindfulness over time through practice can contribute to trait mindfulness \cite{Kiken2015-la}. Mindfulness exercises use therapeutic approaches in both formal and informal practices. Formal practices are more structured mindfulness exercises that can include activities such as sitting and walking meditation \cite{kabat2009full}, whereas informal practices bring the foundations and ideologies (such as being in the present moment) into everyday activities such as folding laundry \cite{hindman2015comparison, terzimehic2019review}. 

While formal mindfulness practice can be effective and helpful to one's well-being \cite{brown2003benefits}, it requires one to find a time and a place to practice. Research has shown that mindfulness can be cultivated not only through rigid meditation practice, but also through everyday mindful activities \cite{Hanley2015-al, Shankland2021-xs}. Informal mindfulness offers more diverse and flexible ways to practice, such as washing dishes \cite{Hanley2015-al} and mindful eating \cite{Shankland2021-xs}, and may be more accessible to novices \cite{Hanley2015-al}. The idea of using technology to help novice practitioners transfer what they learned in mindfulness practice to daily life has been suggested by meditation teachers \cite{lukoff2020ancient}. Within community practices, mindfulness practices are supported both in-person and virtually within teaching and educational sessions for both beginners and more advanced practitioners to learn new techniques and exercises and more community-based experiences where people will meditate together \cite{adkins2010using, el2020design}. In traditional mindfulness practice, community practices are often focused on more formal practices rather than informal ones. 

Mindfulness and meditation are sometimes used interchangeably in HCI research \cite{terzimehic2019review} and in mindfulness research \cite{Van_Dam2018-rk}. We differentiate between the two terms, referring to meditation as a more formal practice with a wide range of techniques that aim to train the mind and achieve a heightened level of awareness or focus \cite{Sparby2021-cd}. Meditation and mindfulness can intersect, for example, through \textit{mindfulness meditation} practice, a practice that involves sitting quietly, focusing on the present moment, and non-judgmentally observing one's breath, thoughts, emotions, and bodily sensations to cultivate mindfulness and self-awareness. But mindfulness and meditation are not the same. Mindfulness, the present-moment awareness with a non-judgmental orientation, can be cultivated through various exercises and activities, one of which being mindfulness meditation. Other mindfulness practices don't necessarily involve formal meditation;  we refer to informal practices as \textit{everyday mindfulness} \cite{Thompson2007-ed}, which involve \textit{"maintaining the open, accepting, present focus of attention during day-to-day life."} 

Moreover, while HCI research has adopted certain definitions of mindfulness, individuals may adapt various definitions based on what they learned, following spiritual or secular traditions. Because mindfulness is highly practice-based, it is also important to understand how each individual translates their definitions into specific forms of practice, and how their definitions and practices evolve over time. By learning about these characteristics and trajectories from experienced practitioners, we hope to inform HCI research and design toward supporting mindfulness practice in the long run. Furthermore, we also note the limited works in HCI that explore both the formal and informal practice of mindfulness \cite{terzimehic2019review} and the difference in definitions between HCI researchers and practitioners. Therefore, in our work, we look to fill the gap by understanding and developing guidelines and prototypes to support mindfulness prototypes not only for the formal practice of mindfulness but also to encourage informal practices. 

We acknowledge the importance of the spiritual and religious elements of mindfulness; in this paper, we focus more on the ways in which practitioners define mindfulness, which may have different aspects, and on the practice of how they \textit{do} mindfulness in their everyday life, as opposed to the more spiritual aspects and their religious achievements.

\subsection{Well-being benefits of mindfulness and individuality}
Scientific findings about the health-related benefits of mindfulness have been well-documented \cite{brown2003benefits}. However, the effectiveness of mindfulness training is affected by individual differences such as personality, dispositional traits, and prior experiences of emotion regulation \cite{Tang2020-ks}. Personality traits, for example, can predict individual preferences for different mindfulness guidance techniques \cite{Tang2020-bc}. Individuals also react differently to different mindfulness practices \cite{Farias2016-nk}. Besides individual differences, research on trajectories of mindfulness practice shows that, over time, novices' mindfulness practice changed to be less effortful, more interesting, and more meaningful \cite{Osin2022-ui}. Research that explored how people first got into mindfulness and what keeps them practicing in the long run, found that practical resources, routines, and support from others help sustain the practice \cite{Birtwell2019-ai}. Such studies emphasize mindfulness practice as an ongoing journey and propose exploring how this change happens over time and how to support this journey to make mindfulness more useful and meaningful to individuals. However, to our knowledge, there is limited work on how technology can be utilized (both technology specifically designed for mindfulness and technology for our means) as a way to emphasize the longitudinal practice of both individual and community-based mindfulness. 

Given this research, one-size-fits-all mindfulness interventions are often considered less effective \cite{Osin2022-ui}. Instead, support for mindfulness should be individually tailored to people traversing through different personalized journeys \cite{Lundh2020-jn}, to increase the receptiveness and outcomes of mindfulness. Individuals may obtain different benefits from mindfulness practice and have different motivations and barriers to the practice. One promising direction is using technology that offers personalization and customization of mindfulness interventions to promote more effective goals and health outcomes \cite{Sliwinski2017-wu}. Yet with this knowledge, it is unclear how future HCI designs can support this understanding of mindfulness. In our work, we look at developing strategies that HCI researchers can use to support the practice of mindfulness. 

Our research does not aim at measuring personalities or traits of mindfulness. Instead, our goal is to explore individualized experiences by obtaining an in-depth insight into mindfulness practitioners' own perspectives on their motivations, experienced benefits, and challenges of keeping consistent mindfulness practice. This investigation provides new understandings and design implications for supporting mindfulness as a meaningful personal journey. 

\subsection{Technology for mindfulness}
HCI research on mindfulness has largely focused on designing, developing, and evaluating interactive technologies that facilitate mindfulness meditation \cite{Vidyarthi2014-yo,Cochrane21, roo2017inner, prpa2018attending, Niksirat2019-yr, Sas2015-bn, Moraveji2012-gs, Wongsuphasawat2012-uu}. Such interactive technologies include online mindfulness classes \cite{krusche2013mindfulness}, wearables \cite{kressbach2018breath, Cochrane2022-zv}, and virtual reality environments \cite{amores2016psychicvr, roo2017inner, prpa2018attending}.  

With a myriad of commercial technologies to support mindfulness (e.g., Calm, Headspace), and research into new modalities such as XR \cite{Dollinger2021-hs}, HCI researchers also examined existing technologies for mindfulness and meditation, the features they offer for meditation practice \cite{dauden2018evaluating}, and how they are perceived \cite{Markum20} and used by meditators \cite{derthick2014understanding, Li2022-beyond} and mindfulness teachers \cite{lukoff2020ancient}. 

However, this group of studies has a preliminary focus on technologies that are used to help with guided meditation and a formal practice \cite{dauden2018evaluating, Dollinger2021-hs} as well as sleep and meditation \cite{lukoff2020ancient}. The people who use these technologies for their meditation practice think about the pros and cons of using them \cite{Markum20, derthick2014understanding}. Only recently, Tan et al's \cite{Tan2023-mindfulmoment} work studied ``on-the-go mindfulness practice'', with the aim to support mindfulness in the context of casual walking with AR glasses. Overall, most technologies were tested over short periods, potentially limiting the examination of long-term use and the benefits of sustained mindfulness practice.

While mindfulness is traditionally viewed as a self-centered and solitary practice, a few HCI and CSCW work explored practicing mindfulness with others via technology such as in video-mediated communition tools, VR and AR technologies. For example, Li and Guo's \cite{Li_Guo_LiveStream} work suggests that people attend live meditation to get the teacher's guidance and to connect with other meditators worldwide. ZenVR \cite{Feinberg2022-lq} mimics in-person meditation learning environment by having virtual pre-recorded teacher voicing the structured curriculum to a group of learners represented by avatars. Besides, Mindful Garden, an AR system, displays guided individual's biosignals as flowers to support co-located mindfulness meditation for greater connection \cite{Liu2022-wz}. These technologies bridge the gap between individual and collective mindfulness experiences and emphasize the potential of incorporating social support in the well-being practice.

In contrast to involving mostly novices in HCI mindfulness research, some research with experienced mindfulness practitioners, those who have been practicing for a long time, found that they use video chat for meditation sessions and to connect with teachers \cite{derthick2014understanding}, and less commercial meditation apps \cite{Markum20}. They are concerned about tracking the practice as an interference with a sense of self \cite{derthick2014understanding}, and about stimulation and distractions coming from technologies that negatively impact transcendent experiences \cite{Markum20}. Recent changes in lifestyles and technology use, especially in response to the COVID-19 pandemic, raise an important question of how experienced practitioners interact with technology for both formal and informal practice have changed. 
According to David Levy's proposal in \textit{Mindful Tech} \cite{Levy2016-jh}, which is based on his own practice and teaching experience, this might resemble mindfulness exercises that a practitioner engages in to become aware of their use of digital technology. This perspective provides a new lens on using technology mindfully instead of outright criticizing its downsides and pushing against it. 

To sum, despite many supporting technologies available, there appears to be a disconnect between the operational definitions of mindfulness embedded in these technologies (short-term, formal meditation, technology support, individual-oriented) and the traditional understanding held by practitioners (long-term, formal and informal practice, mindful technology use, combining individual practice and group/community support). Do experienced mindfulness practitioners welcome technology or stay away from it in their everyday mindfulness practice? In what ways do they use and not use technology? To gain an in-depth understanding of these questions, we carried out our study, presented next. 

\section{Method}
\subsection{Authors' Positionality}
The first and second authors are mindfulness practitioners with seven and twelve years of intermittent mindfulness practice respectively. While we grounded the research in the literature, our backgrounds in mindfulness practice inevitably motivated the research conceptualization, utilization of personal networks for participant recruitment, and interpretation of the data. The third author, not actively engaged in mindfulness meditation, provided an external perspective to enhance the rigor of our data analysis, reflections, and research presentation.  
\subsection{Participants}
Our focus in this research is to understand the practices of people who are beyond the learning stage and have been practicing mindfulness in their everyday lives for a long term. To obtain a nuanced and in-depth perspective, our targeted participants were those who self-identified as regular practitioners of mindfulness. Our inclusion criteria focused on participants' consistency of mindfulness practice over multiple years; we excluded those who are new to mindfulness or who have just ``tried it out''. Starting from the first author's networks, we first approached the mindfulness teachers in the local community and used the snowball sampling strategy subsequently to extend our search for suitable participants. We stopped recruiting participants when we reached data saturation in the analysis.

We recruited 20 individuals who self-identify as regularly practicing mindfulness (13 female, 7 male; ages 18-74; 12 White, 5 Asian, 2 African American, 1 Hispanic; see Table \ref{talble:1}). All participants report more than 4 years of mindfulness practice experience, and twelve of them have mindfulness teaching experience. While previous work distinguished the tradition of participants' meditation practice \cite{lukoff2020ancient, Markum20}, most of our participants didn't identify with any tradition regardless of their religious beliefs, except three participants (Lisa, Grace, Amber) who self-identify as Buddhists.

\begin{table}[htb]
\centering
\caption{Demographics of the participants}
\begin{tabular}{lllllll} 
\hline
\textbf{Pseudonym} & \textbf{Gender} & \textbf{Age} & \textbf{Ethnicity}   & \textbf{\begin{tabular}[c]{@{}l@{}}Years of \\Practice\end{tabular}} & \textbf{\begin{tabular}[c]{@{}l@{}}Frequency \\of Practice\end{tabular} }& \textbf{Form}                                                                                                \\ 
\hline
Alice      & F      & 25-34 & Asian & 5                                                           & 3 times/week                                                    & Meditation                                                                                               \\
Clare      & F      & 25-34 & Asian & 6                                                           & Daily                                                           & Meditation, yoga                                                                                         \\
Paul       & M      & 45-54 & White & 20                                                          & Daily                                                           & Meditation                                                                                               \\
Jasmine   & F      & 45-54 & Asian & 8                                                           & Daily                                                           & Meditation                                                                                               \\
Carson     & M      & 25-34 & White & 7                                                           & Daily                                                           & Meditation, affirmation                                                                                  \\
Karina     & F      & NA  & Black  & 40                                                          & Daily                                                           & Meditation                                                                                               \\
Mike       & M      & 45-54 & White & 20                                                          & Daily                                                           & Meditation, yoga, taichi                                                                                 \\
Alison     & F      & 65-74 & White & 31                                                          & Daily                                                           & Meditation                                                                                               \\
Penny      & F      & 45-54 & White & 7                                                           & Daily                                                           & Meditation, yoga                                                                                         \\
Lisa       & F      & 55-64 & White & 14                                                          & Daily                                                           & Meditation                                                                                               \\
Grace      & F      & 35-44 & Hispanic & 10                                                          & Daily                                                           & Meditation                                                                                  \\
Amber      & F      & 35-44 & Asian & 15                                                          & Daily                                                           & Meditation                                                                                               \\
Steven     & M      & 65-74 & White & 45                                                          & Daily                                                           & Taichi, daily life                                                                                       \\
Cam     & M      & 35-44 & White & 12                                                          & 4-5 times/week                                                  & \begin{tabular}[c]{@{}l@{}}Formal and informal \\ongoing practice\end{tabular}                           \\
Kaylee     & F      & 55-64 & Black & 35                                                          & Daily                                                           & Meditation                                                                                               \\
Erica      & F      & 45-54  & White  & 34                                                          & Daily                                                           & \begin{tabular}[c]{@{}l@{}}Moment to moment awareness, \\sitting and movement\end{tabular}              \\
John       & M      & 55-64 & White & 10                                                          & Daily                                                           & Meditation                                                                                               \\
Shu        & F      & 18-24 & Asian & 4                                                           & Daily                                                           & Meditation                                                                                               \\
Melissa    & F      & 35-44 & White & 20                                                          & Daily                                                           & Yoga, meditation, walking                                                                                \\
Jeremy     & M      & 25-34 & White & 9                                                           & Daily                                                           & Meditation, daily activities                                                                            \\
\hline
\label{talble:1}
\end{tabular}
\end{table}

\subsection{Interview procedure and data analysis}
We conducted interviews from April 2021 to August 2022 via Zoom or in-person, depending on participants' preferences. We developed a semi-structured interview protocol based on the research questions and iterated it following pilot interviews. The interview protocol consisted of four sections: 

\begin{enumerate}
    \item Background: We started by asking participants background questions such as ``how did you get into mindfulness'' and ``how long have you been practicing it''. 
    \item Everyday mindfulness: We then moved to questions about the participant's everyday mindfulness practices, such as when and how they practice, including both formal and informal practice. 
    \item Perceived impacts: We then asked questions about their perceived benefits and barriers to practicing mindfulness. 
    \item Role of technology: Finally, we asked about digital technologies they use to support their mindfulness practices for solo practice or in a group setting and the impacts of these tools. We also encouraged them to show us these technologies. 
\end{enumerate}

Interviews lasted 45-90 minutes, were audio-recorded with participants' permission, and fully transcribed, changing names to pseudonyms. Interview transcripts were imported into Atlas.ti\footnote{\url{https://atlasti.com}}, a qualitative data analysis software. Initial discussions and analysis of the interview transcripts ensued immediately after each interview. Our analysis included open and axial coding \cite{saldana2021coding} with the following steps, repeated iteratively as more data was added from additional interviews: 

\begin{enumerate}
    \item First, two researchers read through the transcripts independently multiple times and conducted an initial open coding. For example, we identified codes such as ``mindful brushing teeth'' and ``mindful washing dishes'' to describe the informal mindfulness activities, ``increasing emotional stability'' and ``reducing chronic pain'' to describe participants' perceived benefits of mindfulness practice, ``using Headspace for guided mindfulness session'' and ``using Zoom to attend group meditation'' to describe technology use for mindfulness. 
    \item Then, the two researchers went over each other's codes, refined the codes, discussed the transcripts together, highlighted excerpts and identified categories into which we grouped the codes (axial coding). For example, we created categories such as ``definitions of mindfulness'', ``strategies for consistent practice'', ``formal and informal mindfulness practice'', and ``technology non-use''. Appendix \ref{table-codes} summarizes the categories we identified and examples of specific codes each category included in the data analysis. 
\end{enumerate}
    
We stopped interviewing participants once we reached saturation and no new themes emerged: all the data collected from our 19th and 20th participants fell into the existing categories and codes we identified in the data analysis process.

\section{Findings}

Based on our analysis, we report findings on how participants define mindfulness practice, characteristics of their daily practice, experienced benefits and challenges, and their use and non-use of technology for everyday mindfulness practice. We organize the findings by our guiding research questions. We start by presenting findings related to the practice of mindfulness regardless of technology, to help contextualize the role of technology, presented later.

\subsection{Everyday mindfulness: Defining, practicing, and finding}
\label{findings - define mindfulness}
\textbf{RQ1: What are the working definitions and characteristics of both informal and formal mindfulness practice from the practitioners' perspectives to give HCI researchers a richer understanding of the holistic mindfulness practice?}
\subsubsection{Defining everyday mindfulness}
We found a range of perspectives on the definition of mindfulness and what counts as mindfulness practice, and these definitions were based on participants' everyday mindfulness activities. One of these definitions, \textit{paying attention to the present moment with awareness}, was often based on formal mindfulness practices, e.g., sitting meditation, walking with awareness, or doing yoga with attention to one's breath. While most participants focused on the concept of awareness in mindfulness, Grace added to her definition the concept of kindness: 
\begin{quote}
  \textit{"How can we be mindful and kind at the same time? [...] Because I think there is a danger of taking mindfulness completely out of its context. Because you could use mindfulness in the military, for your soldiers to be very mindful about how they're shooting their target."} 
\end{quote}

Other definitions were based on more casual activities in participants' daily lives. For example, they defined mindfulness as \textit{doing one thing at a time}, such as focusing on washing dishes. Kaylee explained how eating without doing anything else demonstrates mindfulness: 
\begin{quote}
    \textit{"Let me just use the first bite as a being mindful thing, and then I can talk to my friends. I want people to enjoy their food. I know we're always trying to multitask. But actually, we're not really multitasking, we're toggling between these different activities. And really, the concept behind mindfulness is that it's okay to do one thing at a time."}
\end{quote}

The third way that participants defined mindfulness is as a way of \textit{destressing and relaxation}, such as doing artwork or walking in nature.  For example, Shu treated her Yoga practice before bedtime as \textit{"a process of becoming aware of my body, and then relaxing it. At the same time, it is also beneficial for my sleep"}.

These three definitions were not mutually exclusive for our practitioners' practice. Instead, we found that some practitioners flexibly adopted these definitions and translated them into various mindfulness exercises in their daily lives. For example, Melissa has her dedicated morning yoga and meditation practice and she also practices eating mindfully: \textit{"I started just eating more slowly and without distraction. And that really helped me digest better.[...] I was enjoying the food more too"}. Besides these two activities, she also views tea break as a mindfulness exercise to relax and \textit{"take a break, like in the middle of the morning, make some green tea, and then it's like, very relaxing in it." }. 

The conceptualization and definition of mindfulness can be abstract and complex. However, we found mindfulness practitioners are able to incorporate their own experiences and daily activities to define mindfulness using a bottom-up approach, bringing flexibility and personalization into mindfulness. In other words, they develop an understanding of what mindfulness is based on their own experiences and day-to-day activities of mindfulness, instead of adopting a predefined definition. Understanding this bottom-up approach recognizes the varied ways that individuals engage with mindfulness and helps inform a broad spectrum of designs that are deeply resonant with people's mindfulness journeys. We expand the specific design implications in the discussion \ref{discussion: defining mindfulness}.

\subsubsection{Practicing mindfulness formally and informally}
Participants reported a mix of formal practices and casual activities in their daily life that together comprise their mindfulness practice. They highlighted the importance of their formal practice, associating it with their morning or night routine, as part of their habits. The morning mindfulness routine was particularly valued, as the awareness trained in the morning may be transferred to the rest of the day, as Alison described:

\begin{quote}
     \textit{"I have a sitting meditation at 7 am. That lasts for half an hour. [...] There's the ending bell that stops, "you're done." But the day of my life is a laboratory for exploring how to continue in all inner conditions, all outer conditions. [...]  I want to remember myself as often as possible."}
\end{quote}

Alison's account demonstrates how formal mindfulness practice can carry over to their informal daily activities, especially because, as Grace explained, it is easier to get distracted during day-to-day interactions in complex real-life contexts. Mike also explained,

\begin{quote}
  \textit{"It's harder, because it's more complex. And so because this intention to be attentive is so fragile, one is swept away from it so easily. [...] It does help to set aside a time and place in which there's nothing else that's being tried other than [mindfulness]."}  
\end{quote}

Participants reported that when they just started practicing, their mindfulness activities were more narrow, and focused on sitting meditation or yoga. Over time, they incorporated more mindful activities in their daily lives while keeping the key principle: paying attention to the present moment without judgment. Practicing mindfulness during daily activities and interactions was seen by participants as a way to bring the core of mindfulness into their everyday lives. They mentioned activities such as brushing teeth with full concentration and sensation, using mindfulness for work-life boundary management, and dealing with interpersonal conflicts by pausing to observe physical sensations without judgment and immediate reactions. Karina said: 
\begin{quote}
    \textit{“I meditate every day, but I also just do things mindfully. So if I'm doing the dishes, I'm paying attention to doing the dishes. I'm not watching a television show and talking and having a conversation at the same time. Because that's not mindfulness.”}
\end{quote}

Further, despite the importance of formal practice, participants mentioned that when life is going well, or when they are getting busy, formal practice may not be prioritized, but they still tried to keep the habit of mindfulness through informal practice. As experienced practitioners, casual mindfulness practices such as mindful eating, walking, and exercising, were easier to integrate into their daily lives. Unlike formal mindfulness practices such as sitting meditation, informal mindfulness activities didn't require them to find a specific time and place to practice. For example, Jeremy, a barista who drives to work every day and washes dishes as part of his job, incorporates these two activities in his mindfulness practice. Mindful driving allows him to be awake for safety and dishwashing permits him to pause the work to \textit{"settle myself from the hecticness of everything"}. He described the importance of these moments for \textit{"recollecting myself, [otherwise] I would probably be a much less functional person.}"

\subsubsection{Finding the appropriate form of practice}
\label{findings: Finding the appropriate form}
As experienced mindfulness practitioners, our participants reflected on choosing suitable activities to practice mindfulness in their daily lives. With various formal practices available such as observing one's breath, body scan, and mindfulness movement, they expressed that these meditation practices may not be suitable for everyone. As the core of these practices is to cultivate non-judgemental awareness, it is acceptable to experiment with different types of practice and identify what one mostly resonates with and can maintain in the long run. For example, Kaylee, a mindfulness instructor, mentioned that in her teaching experience, she encountered people who were not able to do sitting meditation: 

\begin{quote}
    \textit{"We ask them, have you had serious addictions, trauma, major depression, it may be that a sitting meditation is not the right thing. And especially anxious people can't sit for half an hour. [...] So in that case, uh, moving, a walking meditation, mindful yoga. Something that is connected to the body in motion is hugely helpful."}
\end{quote}

Paul also proposed that beginners start with a \textit{dynamic form} of practice to help understand the nature of mindfulness: 

\begin{quote}
    \textit{"Something else that's immediate, and you don't have to learn anything, is the walking meditation. Go for a walk, and for ten minutes, try the exercise of turning off thinking, keep senses on what you feel, hear, see, smell, keep your thoughts to what you're experiencing in the present moment."}
\end{quote}

We found that while practitioners' formal practices were \textit{standardized} and \textit{structured}, based on what they learned through in-person mindfulness programs or online resources, informal practices were more \textit{flexible} and \textit{personalized}, given one's daily routines. Repetitive daily chores such as sweeping the floor and washing dishes were in particular seen as opportunities for informal practice. Therefore, the social aspect of mindfulness is more easily identified in the formal process rather than the informal process of mindfulness. Mike explained \textit{"It helps if there's a simple physical task, it can be repetitive, or just something simple, physical, that the attention can stay with the body as it does."} These tasks do not require excessive mental resources, and involve multiple bodily sensations such as touch, hearing, sight, smell, and taste that can be attended to while doing the task. Transforming these tasks into a mindfulness activity helps train awareness of being present and relieves the boredom of doing the task. Jasmine explained: 
\begin{quote}
    \textit{"I get irritated with, household chores are so boring, like never-ending but [...] focus on the hot water, focus on the sensation of the soap, feel the vessel with your hand that itself is a meditation, do the dishes can be. [...] focus on the actual act of doing the dishes. And that helps me have a better spiritual about chores."}
\end{quote}

%Similar to Jasmine's account, other participants also favored activities that involve multiple bodily sensations such as touch, hearing, sight, smell, and taste. For example, during the interview, Alison described her experience of feeling her feet and hands, \textit{"my feet are not on the floor. I could maintain that sensation, [...] or even be aware of one or both hands, what they feel like when I hold the cup."}
%\end{quote}

However, participants commented that practicing everyday mindfulness in real life can be distracting, especially for novices, and it may be easier to start with one or two informal exercises and expand the practice once one becomes skilled. Grace explained:
\begin{quote}
    \textit{"Every activity can be an opportunity, however, for beginners it's probably better to start with certain conditions and not start, say in the middle of a baseball game, sitting in the crowd. Ultimately, [...] when you're more experienced, then you can do it anywhere."}
\end{quote}

Finally, informal practice can be chosen based on one's purpose for practicing mindfulness, such as sleeping better, reducing stress, or improving one's relationships. For example, several participants mentioned that to improve relationships, one could practice mindful listening, by fully engaging in a conversation, observing body sensations, pausing before responding, and then making an informed response instead of unconsciously reacting. Such informal exercise can be very useful but is also difficult to apply as it requires one to face their stressor and self-observe before reacting. However, supporting such informal mindfulness practice makes one's mindfulness practice more meaningful. For example, Mike, a mindfulness instructor, reported that his students come to him to learn mindfulness practice to solve life issues (e.g., stress, divorce, loss of loved ones), and he tries to tailor the practice to their needs outside of meditation. He explained that the degree to which they feel less stressed, more empowered, and more creative in these daily situations is a meaningful measure of the progress of their mindfulness practice.

\subsection{Benefits, barriers, and strategies of maintaining practice}
\label{findings - benefits and barriers}
\textbf{RQ2: What are motivations, perceived benefits, and challenges faced by mindfulness practitioners in maintaining a regular practice? How can we learn from these insights to inform designs that support consistent mindfulness practice?}
\subsubsection{Benefits and motivations}
\label{findings - vitamin}
Our participants reported short- and long-term physical and mental benefits of mindfulness, which motivated them to continue practicing it consistently. %Carson said: 

In the short-term, participants reported immediate benefits right after a short mindfulness session, such as a 10-minute sitting meditation or a half-hour mindful walk: feeling calm, relaxed, and clear-headed, and getting better sleep. Carson said: \textit{"if you have a good mindfulness session or practice, then you feel relaxed and centered."} Melissa described that a practice session \textit{"lifts a veil."}

Several participants saw the pursuit of short-term benefits as stepping stones to long-term practice. But some also cautioned that the short-term benefits may just be the byproducts of the cultivation of non-judgmental awareness and these effects vary for different people. They saw the real benefits of mindfulness experienced in the long-term process of self-discovery. Steve explained that it is possible to attract people to mindfulness through short-term benefits, which he called "extrinsic", but they will need to find the balance with long-term "intrinsic" benefits to maintain the practice, as there is no \textit{"finish line"} in mindfulness:
\begin{quote}
    \textit{"It's extrinsic benefits and intrinsic benefits. And in anything that we do, there are some benefits that are very obvious, very external, very measurable, like a finish line. Then there are the intrinsic benefits, which are much deeper, much more subtle, but much more profound."}
\end{quote}

The long-term benefits participants reported included enhanced emotional regulation, reduced chronic pain, increased resilience and a sense of control over their life, and improved interpersonal relationships. These experienced benefits motivated participants to keep practicing consistently. Some participants even attributed their well-being to the practice of mindfulness, attributed messiness in their daily lives to the lack of mindfulness practice, and felt that their life would have been worse if they had not kept the practice. Paul explained: 
\begin{quote}
    \textit{“Whatever happens with my health down the road, I know that it won't be as severe as it would be if I wasn't doing yoga, if I wasn't doing meditation, if I wasn't doing those things. Maybe I would be taking antidepressant pills.” }
\end{quote}

Overall, while some of the reported benefits of mindfulness align with previous findings \cite{brown2003benefits, Creswell2017-sk, Grossman2004-vf}, we identified two metaphors that participants used to relate to the benefits of mindfulness: \textit{medicine} vs. \textit{vitamin}. This perspective could be especially insightful for designers aiming to promote consistent mindfulness practice for mental well-being, as we further elaborated in the discussion section \ref{design through metaphor}. Some participants, for example, Paul, referred to mindfulness as a \textit{medicine}, practicing it to obtain mental and physical health, as Melissa said \textit{"If I do those practices, it just keeps me grounded. I just feel like everything's okay."} Others related to mindfulness as a \textit{vitamin}, where it is a habit and has preventative effects on one's well-being in the long term. These two metaphors may co-exist for practitioners, who practice mindfulness both as medicine to respond to immediate difficulties in their lives and as a vitamin to maintain their long-term well-being. 

\subsubsection{Barriers}
Despite the benefits, all participants, even those who have established a long-term daily practice routine, felt that sticking to the practice was not easy. Steve described being able to maintain the practice as a privilege: \textit{"I feel that it is a privilege to live the life that I do, because I have the opportunity each day that I am training, to choose a healthy relationship and empowering relationship with stress".} 

Busy or inflexible schedules were frequently mentioned as a barrier to keeping the practice. Some participants, especially those who share a workspace with others or who take care of kids, found it difficult to fit into their schedules formal mindfulness practice, which requires a quiet and undisturbed space. Kaylee described that when her kids were young, \textit{"I would always get up about an hour before my kids would get up, so that I would have the time to sit and to come into the present moment."} Busy lifestyles also led to feeling too tired to practice, like Shu, who had to fight her own resistance to practice after \textit{"I have been in front of the computer for a long time, I would then feel very tired physically and mentally"}.

Sometimes it wasn't about finding a time or a place to practice, but about experiencing strong emotions or physical pain, or a fear of facing these uncomfortable sensations during formal practice. Jasmine eloquently described the conflict between forcing herself to keep the practice regardless of how she felt in the moment and the core of mindfulness: being non-judgmentally aware of the present moment: 
\begin{quote}
    \textit{"I sometimes feel very tired. Or I'm not feeling well, my pain is some days worse. I don't feel like being mindful. And on those days, in a weird way, it is mindful in the sense that I know exactly what I'm doing." }
\end{quote} 

Another difficulty participants faced was related to breaking habitual patterns. Creating a habit of mindfulness may be especially difficult as mind-wandering seems to be the default mode of the human brain \cite{Killingsworth2010-ky, Buckner2008-qb}. Some of our participants who were teaching mindfulness observed this in their students who are new to mindfulness, as Erica said: \textit{"there may be resistance because this is a new practice. And for all of us changing habits, it's always the hardest, but give it a try."} Even the very experienced practitioners faced their own resistance after years of practice. Being aware of and normalizing the difficulty, they still sat down on the cushion for the practice, as Alison, with 30 years of experience, said:

\begin{quote}
    \textit{"The resistance is never going to go away. You just have to learn. It's like my teacher told me about some other things. It's just like a stick in the road. You hop over it. You go around it."}
\end{quote}

\subsubsection{Strategies}
Although not every practitioner encounters the same difficulties in mindfulness practice, we identified three primary strategies that our participants adopted to maintain a long-term mindfulness practice: \textit{practicing short mindfulness exercises, leveraging social accountability, and seeking help from a teacher. } We further discuss the practical design implications of these strategies for supporting mindfulness in section \ref{implication: bringing community support}.

Practicing short mindfulness exercises was a commonly used strategy among our participants to maintain their practice. The short practice might be part of morning or night routines or during brief breaks throughout the day, as Erica suggested, \textit{"even use the phone alarm to pause five or 10 times a day, to pause to come into the actuality of our living over a moment"}. Maintaining mindfulness doesn't mean a certain number of minutes of daily practice; showing up for the practice is more important in the long run. Jasmine explained: 
\begin{quote}
    \textit{"It doesn't matter how long you do it, or how well you do it. What matters is you show up every day. So some days, when I had a lot of pain, I couldn't do more than 10 minutes. So I'd said it's okay. But I still counted it as a practice because of letting go of needing to be perfect." }
\end{quote}

Although participants mostly practice mindfulness alone, being part of a community makes them more accountable for the practice. Such communities can be in different forms: in-person mindfulness programs, Zoom weekly meditation groups, and family and friends around them who also practice mindfulness. Katrina said that her practice was stronger \textit{"whenever I'm in a community."} For Cam, mindfulness was \textit{"more of like a social thing as opposed to just like doing it individually on my own. I was having weekly meditation group and having a community, or speaking with my wife or kids about it, where there was some level of accountability and reminders socially about it"}. 

Seeking help from, and the presence of a teacher was another strategy that participants reported helped them discover and learn about mindfulness, answer questions along with the practice, and overcome barriers along their mindfulness journeys. Some of our participants who were mindfulness teachers talked about helping their students. Mike, for example, helped beginners get into mindfulness by shortening the practice, offering guided meditation recordings, and working with his students to identify possible times to practice, \textit{"and then observe and see how it goes and modify as needed."}. 

Digital technologies also play an important role in supporting mindfulness, as presented next.

\subsection{Technology use and non-use for mindfulness practice}
\label{findings - tech use and non-use}
Our third research question asks \textbf{"What roles do technologies play in practitioners' everyday mindfulness practice? How does it inform the design of future social computing technologies for mindfulness and mental well-being?"} We present the findings related to the first aspect of the question below and elaborate the latter part more thoroughly in the discussion section. 
\subsubsection{Using technology to remind, assist, and record}
 
Table \ref{table:2} presents a list of the digital technologies that our participants reported using or have tried for their mindfulness practice, some of which have been directly designed for this purpose. For example, participants mentioned using mobile meditation apps such as \textit{Headspace} and \textit{Calm} that provide guided resources and helped create an atmosphere of mindfulness practice with ambient music, \textit{Eat Right Now} for learning mindful eating, and online journaling apps such as the Japanese \textit{Awarefy} for recording emotions. 

\begin{table*}[htb]
\centering
\caption{Summary of digital tools reported by experienced mindfulness practitioners and primary uses for mindfulness practice}
\begin{tabular}{llll}
\hline
\textbf{Technology} & \textbf{Ways used for mindfulness practice}                                                                                                                    \\ 
\hline
Headspace              & Guided meditation resources                                                                                                                                \\
Calm                   & Guided meditation resources; ambiant music                                                                                                                  \\
Insight Timer          & Guided meditation resources; Timer; Community features; Record length of practice sessions                                                                      \\
Awarefy                & Check in and record emotions; Self-care activity recommendations                                                                                                \\
Liberate               & Guided meditation resources with voices from people of color                                                                                                \\
Eat Right Now    & Short lessons and guided mindfulness resources for mindful eating                                                                                           \\
Waking Up        & Streamlined lessons to relearn mindfulness theory and practice                                                                                              \\
Meditation Timer & Set duration; Set start, interval and end sounds for a practice                                                                                             \\
Audio files      & Guided recordings from teachers in offline classes                                                                                                          \\
Spotify                & Meditation playlist; Ambient music                                                                                                                          \\
Youtube                & Mindfulness video resources                                                                                                                                       \\
Zoom                   & Online mindfulness group sessions   
\\
Calender               & Reminders and blocked times for meditation                                                                                                                                                                                                                          \\
Google sheets          & \begin{tabular}[c]{@{}l@{}}Record daily life events, emotions and mindfulness practice. \\Find correlation of practice and emotion\end{tabular}  \\
\hline
\label{table:2}
\end{tabular}
\end{table*}

Our participants appreciated the easy access to lots of mindfulness resources with these apps, as Grace described \textit{"it's democratizing access to something that perhaps is harder to get access to."} In addition to using these resources for their own practice, some of our participants recommend appropriate resources from these platforms to others. For example, Erica tried resources on Insight Timer before recommending to her students: 
\begin{quote}
    \textit{"I like to try a practice before I recommend it to a student or anybody. So I've recommended things for sleep, like a guided meditation for sleep, or for just anxiety and panic, or just different things. The one that I've used the most is Kristin Neff's five-minute self-compassion break."}
\end{quote}

Choosing the right resources for different scenarios requires prior mindfulness experience, and getting too many resources might be overwhelming. To deliver mindfulness in a simple and clear way, Cam recommended the app \textit{Waking Up} to his friends for its streamlined lesson and exercise design. In addition to actively using guided resources for daily meditation practice, participants used these digital applications for re-establishing meditation habits, especially after a long period without practice. For example, Cam relearned the same content on \textit{Waking Up} but still got benefits from it, \textit{"even if I knew the content, just the accountability, and the structure for continuing to practice would be important, even if it was repeated content"}. Likewise, Amber uses this app to create a guided structure for her meditation practice after she left a temple where she had a very structured practice. 

Aside from using dedicated mindfulness technologies, some participants use other technologies that apply mindfulness principles in specific life situations. For example, Alison used to have an eating addiction and she is aware of how mindfulness could help: \textit{"If I'm feeling bad, and I reach for a piece of chocolate, and then it makes me feel better. Something happens there, under the layers, and I want more. I just want more. And this is how addictions get rolling. So with mindfulness, we're asked to pause."} The mindful eating weight loss app \textit{Eat Right Now} helped her apply mindfulness to her eating addiction, \textit{"I tried it for a month and it made a shortcut to things I have spent years learning." }

At the same time, participants were aware of some aspects of these applications that presented a conflict with the principles of mindfulness. For example, some of the apps apply social or reward mechanisms to motivate keeping the practice. While the original purpose may be to leverage social accountability, it might induce social comparison, which Cam found unhelpful: \textit{"I feel Headspace is bad at this, where they gamify it and it becomes a social thing, or they encourage streaks or communication in a way that's more hierarchical and competitive."} Similarly, \textit{Insight Timer} has the feature of recording the length of the daily practice and how many consecutive days one has practiced. This design motivated Carson to keep consistent practice, but also presented a conflict about losing the internal motivation for practicing mindfulness: 
\begin{quote}
    \textit{"I'm kind of conflicted about the stars, I mean I do feel better after I meditate. I generally do like having a daily practice, right? But creating these kinds of incentives from the technology to get your stars or keep your streak is bad because it's somehow like an external motivation, and it should be more internally driven."}
\end{quote} 

\label{findings - non-mindfulness tools}
Interestingly, our participants also discussed a range of digital technologies that aren't specifically designed for mindfulness practice. These included productivity tools such as an online calendar and spreadsheets. For example, some scheduled mindfulness practice sessions in their calendars to remind them to practice. Reminders were set at a fixed time or multiple times throughout the day as opportunities to take a pause. This was similar to reminder features in some mindfulness apps that send notifications to practice mindfulness. While reminders were generally considered helpful, Amber cautioned that reminders should be used to scaffold building a habit but eventually one has to develop the ability to remember to be mindful on their own: 
\begin{quote}
    \textit{"Some apps have alarms that can remind you, now it's time to do 30 seconds of meditation. Now it's time to be mindful and there's something about having something tell you to be mindful. That, in the beginning, is really helpful but then I think there becomes a certain point where if you're not learning to do that on your own, then are you really learning mindfulness? Part of mindfulness is you engaging with your mind and remembering to do that throughout your day."}
\end{quote}

Besides setting reminders, participants also used technology to record their mindfulness journey. Jasmine used \textit{Google Sheets} to record daily events, emotions, and mindfulness practice, then used calculations to find relationships between her practice and emotions: 
\begin{quote}
    \textit{"I try to log down at night. I try to be very ruthless with myself about that. [...] I made a bar graph with the number of times I got angry, it was direct, it was inversely correlated with the number of times that I did meditation. So it showed that the less meditation I did, the more angry I got."}
\end{quote}

Participants also used music and video platforms that weren't designed for mindfulness practice to find guided resources and ambient music for their practice, or sounds of bells to signal the beginning and end of meditation. Unlike meditation apps that required spending money or time to learn the app, \textit{Youtube} and \textit{Spotify} were already integrated into participants' daily use for other purposes, and the transition to using them for mindfulness practice seemed natural. Clare described the convenience of using \textit{Spotify} for guided meditation:  
\begin{quote}
    \textit{"I just discovered that some teachers upload their audio recordings to Spotify, and they make it as a playlist. So I think that's very convenient for me, because I use Spotify everyday to listen to music. And I can also use that for my sleeping meditation or guided meditation, even though the resources are not as much as Insight Timer and Calm, it's convenient enough for me to open the app and start doing it."}
\end{quote}

\subsubsection{Using technology to connect with the mindfulness community}
\label{findings - use technology for community}
In addition to these resources, participants also connected with their mindfulness community through digital technologies such as \textit{Zoom}, instant messaging tools, or even recorded talks of their teachers. %As discussed earlier, 
Connecting with others creates a sense of community and helps practitioners to maintain their practice in the long run. For example, Erica discussed the potential of online groups to facilitate one's ongoing mindfulness journey through mentor and peer support.

The COVID-19 pandemic specifically opened the door for technologies such as \textit{Zoom} for online group mindfulness sessions. Penny felt that \textit{Zoom} relieved her anxiety about teaching meditation and yoga in-person during the pandemic, \textit{"I'm in home space, not a lot of people around me"}. Online mindfulness sessions weren't considered ideal compared to in-person for human connection and environmental cues. At the same time, they allowed flexibility of time and place for instructors and students, and had the benefit of controlling the camera and microphone during the session. Jasmine explained:
\begin{quote}
    \textit{"I have allergies and often sneeze. So I can just mute it. I can sneeze and it doesn't disturb. [...] But if I sneeze during a guided session in person, that was awkward for me."}
\end{quote}

Comparing \textit{Zoom} sessions with listening to guided recordings, Paul said that \textit{"there are real people and there's some community like-mindedness."}. In other words, live online group meditation sessions brought a sense of social presence and connectedness that recordings didn't, when in-person connections were not an option at the height of the pandemic.   

\subsubsection{Using technology as an object for mindfulness}
\label{findings - mindfulness as an object}
As technology has become an indispensable part of everyday life, in a few cases, participants applied mindfulness as part of their technology use habits. In his mindfulness teaching, Mike encouraged his students to turn checking their phones into an awareness exercise, \textit{"for many people, checking email or social networks or automatically checking the phone is a habit that they've been aware of and are interested in becoming more aware of."} This echoes to mindful information exercise introduced in \textit{Mindful tech} \cite{Levy2016-jh}.

Alison associates technology use with mindfulness practice by doing specific mindfulness exercises of checking her emotions, applying breathing exercises, being aware of the intention of use before looking at her phone or starting to work on the computer:
\begin{quote}
    \textit{"There's a whole set of habits and patterns that are happening in our relationship to technology, that can be the object of mindfulness practice. So I can start to become aware of my automatic looking at the phone, or I can start to be aware of my emotional state, as I check to see if a post was liked. [...] Every time you pick up your phone you can associate it with being very present for three breaths, something like that. Every time I start a new project on a computer, or if I sit down and touch the keyboard to not make it external to your practice, but to incorporate everything."}
\end{quote}

In this way, instead of viewing technology as a separate means that delivers mindfulness or provides external reminders, the very act of using one's devices turned into an opportunity to bring mindfulness to the present moment, and facilitate meaningful digital devices use.

\subsubsection{Technology non-use to defend mindfulness}
\label{findings - tech nonuse}
Despite the benefits of technology for mindfulness practice, our participants also described the downsides of technology, like relying on built-in incentives to practice mindfulness, as discussed above. They also expressed concerns about over-reliance on phones in particular to practice mindfulness, as Jasmine expressed: \textit{"If someone says, Oh, I don't have a smartphone, so I can't meditate. That should not become an excuse."} Jeremy expressed a conflict between using the phone for mindfulness resources and the costs that come with phone overuse:
\begin{quote}
\textit{"There are great databases of free meditations now. You just have to download an app. And then you're on your phone again. And like Insight Timer, it's a great app. It has great meditations on it. I like doing them sometimes. But then I'm on an app. And then once I finish, I have to open my phone to turn it off. And then I'm on Instagram again. "} 
\end{quote}

Many participants mentioned that sometimes \textit{removing technology} helps defend mindfulness, by providing a mindful break from screens and eliminating distractions. For example, silencing incoming messages on one's digital devices helped people stay more focused during the practice by creating an unobtrusive offline environment and reducing participants' anxiety about receiving notifications. Melissa explained: 
\begin{quote}
    \textit{"I never have my phone on. It's always on silent, always. I don't even put it on vibrate, like just silent. [...] It's so distracting. [...] If I have it on, I feel distracted, even if there's no sound and no one's messaging me, I feel like someone's going to message me and the thought of being interrupted, makes me stressed. So I just keep my phone off."}
\end{quote}

Sometimes, participants had limited success with using built-in features to defend themselves from mindlessly using their phones. For example, Jeremy applied a black and white mode to his screen to make it less attractive, but he \textit{"just looks at the thing in black and white. It doesn't change much."} Grace used a 20-minute timer to limit her social media use, but when the time is up \textit{"I override and I click and more and more."} She imagined that instead of applying device built-in features, a mindfulness exercise could help overcome and replace the mindless social media use:
\begin{quote}
    \textit{"What if there was a meditation practice, a five-minute pause of like, okay, just breathe and then you cannot go back in the thing. You're going to feel anxious and upset that you're not able to, but a guided meditation, ticking with your body to keep with your breathing. Go out for a walk, stretch your body, things like that."}
\end{quote}

Finally, some participants felt that they needed to completely remove their devices from their environment, especially when they were with other people. This helped them be more focused on their interactions with others, as Penny tried to have family dinner time without looking at screens, \textit{"we try at least once a week on Friday to sit together and eat and no phones."}. For Melissa, putting her phone away when she was with other people helped her establish a healthy boundary of phone use:  \textit{"If we have people over, I don't even look at my phone, I just put it aside."}

To sum, our participants were aware of the advantages and disadvantages of using technologies in their mindfulness practice and in their everyday lives. As experienced practitioners, they were thinking carefully about which technologies would be useful and how to incorporate and adapt these technologies in their practice. At the same time, similar to previous research \cite{Markum20}, our participants were also mindful about when these technologies become a hindrance and should be put away or turned off, and came up with ways to \textit{undesign} \cite{pierce2012undesigning} distracting technologies from their mindfulness lifestyle.

\section{Discussion}
This study examines mindfulness practitioners' everyday practices, the benefits and barriers they experience, and technologies they use and don't use in supporting mindfulness practice. In this section, we first discuss what we learned about mindfulness and how we expand the definition of mindfulness in HCI. Then we discuss how our study findings extend the current HCI and CSCW research and provide recommendations for designing future mindfulness tools for mental well-being. 

\subsection{Expanding the definition of mindfulness} %to support a personally meaningful journey}
\label{discussion: defining mindfulness}
Our findings (\ref{findings - define mindfulness}) presented an in-depth understanding of mindfulness as an ongoing personal journey, involving various definitions, forms of practice, experienced benefits and barriers, and the interactions between these factors. The mindfulness literature often defines mindfulness as a non-judgemental present-moment awareness \cite{kabat2009full, Bishop2004-un}. Research in HCI defines mindfulness in a similar way, often focusing on the formality of the practice in the definition \cite{terzimehic2019review}. However, we found that experienced practitioners perceive mindfulness as a multifaceted concept that goes beyond this conventional definition, encompassing a range of informal practices in daily life. 

Consistent with the debate on how to define and operationalize mindfulness \cite{terzimehic2019review, Van_Dam2018-rk}, we found that mindfulness practitioners have a range of definitions of what consists of mindfulness practice. In addition to the common definition of \textit{``paying attention to the present moment without judgment''}, our participants defined mindfulness as the ability to focus on one activity at a time, as well as destressing and relaxing from everyday pressures. 

These experienced practitioners found suitable formal and informal mindfulness practices based on their personal goals, daily activities, and life situations. Informal mindfulness practices were often more accessible and easier to incorporate into daily life, such as eating, driving, doing house chores, and interacting with others. This expanded understanding of mindfulness in HCI, with both formal and informal practices, emphasizes the importance of considering mindfulness as a long-term practice and lifestyle. Although this perspective aligns with findings from some prior mindfulness literature \cite{hanh1999miracle, Shankland2021-xs, Fredrickson2019-dk}, HCI work has mainly focused on the formal practice to define mindfulness.

Our findings suggest that considering the range of definitions of what counts as mindfulness from practitioners' perspectives can enrich the design space for mindfulness technology. We call researchers to broaden the definition of mindfulness by taking into account the dynamic nature of the practice, the various forms it can take, and the individual experiences and preferences of practitioners.

\subsection{Tension between supporting short- and long-term mindfulness practice}
A key challenge in designing mindfulness technologies lies in balancing the support for both short-term and long-term mindfulness practice. Our findings (\ref{findings - benefits and barriers}) indicated that, despite the benefits mindfulness brings, even experienced practitioners who have established a long-term daily practice routine felt that sticking to the practice was not easy. 

The tension between short-term and long-term benefits in mindfulness practice highlights an intriguing point, specifically the contrast between Western and Eastern perspectives. Eastern mindfulness traditions typically emphasize it as a lifestyle, while Western approaches often focus on stress relief \cite{schmidt2011mindfulness}. This discrepancy can be discouraging for practitioners trained in Eastern methods, who perceive mindfulness as a comprehensive lifestyle transformation rather than just a tool for health benefits. However, when designing for a Western audience, it is essential to acknowledge the short-term benefits as an initial step to capture users' interest and inspire them to delve deeper into mindfulness practice \cite{aronson2004buddhist}.

By emphasizing short-term benefits as incentives for behavioral change \cite{Fogg2002}, mindfulness technology can inspire beginners to adopt mindfulness consistently and practically in their lives and facilitate the habit of building long-term practice. Incorporating brief, easily accessible mindfulness activities is one opportunity to explore, following a growing body of research that highlights the benefits of micro breaks and short pauses on well-being \cite{Zacher2014-ty} and happiness \cite{Kaur2020-qz}. 

By designing technology with a focus on promoting these small, manageable exercises, we can create a more approachable and engaging mindfulness experience \cite{lukoff2020ancient}. For example, mindfulness technology that emphasizes post-practice feelings and incremental health benefits can help support beginners initially and gradually shift their focus towards appreciating the present moment, with open and non-judgemental awareness. 

Supporting an ongoing mindfulness journey also means providing people with meaningful motivation to keep consistent practice. Reward mechanisms in mindfulness and wellness systems need to be personally meaningful and intrinsic \cite{Weiser2015-zy, specker2021digital}. A mindfulness tool that counts time one practiced may be focusing on what Baumer \& Silberman \cite{Baumer2011} call \textit{computational transformation}, rather than addressing the intent of mindfulness practice: attention, reflection, and awareness. Instead of, for example, displaying the number of minutes and days of an individual's practice, a system can rephrase it as how many minutes of mindfulness a person has contributed to the community. Alternatively, transforming numerical data into visually expressive formats, such as coloring a mandala \cite{Tucci2001-mandala} to represent days of practice, may offer rewards in a more artistic and personally meaningful form. At the same time, designers need to be cautious about social comparisons that may pressure users to feel an obligation to practice, as has been emphasized in previous work \cite{lukoff2020ancient, Laurie2016-ud}.

Besides supporting motivations for novice practitioners, it is essential to consider the needs and preferences of more experienced practitioners. Our findings (\ref{findings - tech use and non-use}) suggested long-term mindfulness practitioners often have a different approach to technology than newcomers, viewing mindfulness as a lifestyle rather than merely a tool for stress relief. Their use of technology related to mindfulness tends to be selective and purposeful, relying less on guided practices and more on traditional tools or apps that indirectly support mindfulness, such as scheduling apps, timers, or music and video platforms. Focusing on adding reflection, adaptability, and personalization to those existing technologies can help ensure that technology remains a useful and supportive tool for those who have incorporated mindfulness into their daily lives. HCI researchers have looked at both how blenders can be mindful in everyday contexts \cite{van2016engagement} and more traditional practices and how technologies can bring mindfulness communities together using platforms such as YouTube \cite{schmidt2011mindfulness}.  
Our research highlights the significance of everyday digital technology for mindfulness; participants mentioned the importance of digital calendars and spreadsheets in their practice. We propose future mindfulness interventions for more advanced practitioners might want to instead of just having in-app reminders   

The tension between demonstrating immediate effectiveness and addressing the ongoing nature of mindfulness practice presents a critical consideration for designers. While previous HCI work focuses on evaluating the effectiveness of mindfulness technologies for a short period of time \cite{dauden2020body, Terzimehic2022-xh}, it remains unclear how these technologies would impact well-being and other life aspects in the long term. The benefits and depth of mindfulness practice increase with time and consistent engagement; therefore, testing and evaluating mindfulness technologies over the long term is crucial instead of solely assessing their short-term impact or acceptance. This is consistent with calls to expand HCI research beyond the short-term focus on efficiency and productivity toward exploring the design and integration of slow, long-term technology in everyday life \cite{Odom2014-so,Odom2018-md, Odom2019-yt}. In the context of mindfulness technology, longitudinal research and design efforts would allow for gaining a comprehensive nuanced understanding of its long-term impact. Designers could consider how the technology supports the users' mindfulness journey, from beginners who might initially be drawn to short-term benefits to experienced practitioners who seek to deepen their practice over time. For example, using an autobiographical design approach \cite{Neustaedter2012-lr, Cochrane2022-zv, Li_Guo_LiveStream} for mindfulness products may allow researchers with mindfulness experiences to design and use the technology themselves, and gain valuable insights into the design process and user experience.

\subsection{Design recommendations}
\subsubsection{Designing for mindfulness through the lens of metaphor}
\label{design through metaphor}
The use of metaphor to \textit{``understand one thing by describing it as though it were another''} \cite{Blackwell2006-rp} as a design tool has a long history in HCI. The use of metaphors can provide valuable insights into designing technology for mindfulness and affect specific design decisions. Our findings (\ref{findings - vitamin}) suggest two metaphors for the roles that mindfulness play in participants’ daily lives: \textit{vitamin} and \textit{medicine}. Using the ``vitamin'' metaphor for mindfulness implies the need for ongoing support to integrate mindfulness into daily life, similar to taking daily vitamins to support ongoing health in the long run. Design features may include scheduling, planning, contextual reminders for informal activities, and mobility to ensure mindfulness practices are accessible wherever users may be. For example, the Headspace app includes a feature that recommends a daily itinerary of mindfulness activities, from morning routines to evening wind-downs, which follows the ``vitamin'' approach by promoting regular, sustained engagement with mindfulness practices. The ``streak'' feature, popular in mindfulness apps, also resonates with the ``vitamin'' metaphor.

On the other hand, if mindfulness is perceived through the ``medicine'' metaphor, the design should focus on detecting stressful situations or events that require intervention, similar to taking painkillers during moments of pain or discomfort. This could involve monitoring a user's heart rate or sleep patterns, and prompting the system to suggest mindfulness practices when necessary.

The specific design features incorporated into a product will be influenced by the metaphor through which mindfulness is perceived, either as a ``vitamin'' for ongoing support or as ``medicine'' for an intervention during stressful times. A third approach is to integrate both ``vitamins'' and ``medicine'' metaphors in mindfulness design. For instance, many wearables already cater to short-term needs, such as watches that remind users to breathe during stressful times, serving as ``medicine''. Over time, the use of a wearable designed for mindfulness may gradually switch toward a ``vitamin'' function, such as more assistance in recognizing everyday activities, like dishwashing, commuting \cite{Tang2020-bc}, as opportunities to practice mindfulness.

\subsubsection{Using a toolkit approach to design for mindfulness}
Our findings (\ref{findings: Finding the appropriate form}) indicated that experienced practitioners experimented with different forms of mindfulness practice and over time found the appropriate practice based on their changing needs. At the same time, most mindfulness technologies focus on a small number of mindfulness practices, such as meditation or breathing exercises. Instead of prescribing a one-size-fits-all solution, we suggest adopting a toolkit approach, offering various mindfulness techniques, exercises, and activities to cater to different preferences in different situations. A toolkit approach acknowledges that individuals have unique needs and ways of engaging with mindfulness practices. 
Moving beyond supporting breathing or meditation exercises, it is important to consider alternative everyday mindfulness activities that focus on awareness of other sensations, such as mindful eating, walking, and cooking, to accommodate a wide range of changing preferences and abilities. A toolkit to support everyday mindfulness activities could encourage individuals to explore various mindfulness practices to determine which methods are most effective for them at any point. One example is Chen et al.'s ColorAway, with features for recollecting and evaluating one's travel experiences toward gaining self-knowledge through interacting with their travel photos \cite{Chen2018-gz}.

Using a toolkit approach does not focus on providing a single, specific mindfulness exercise. Instead, it offers a meta-method for people to explore, create, and experiment with various mindfulness activities, ultimately discovering the ones that work best for them at different stages of their mindfulness journey. This shift in focus has the potential to lead to innovative and creative ways of seamlessly integrating mindfulness into everyday activities, fostering a more personalized and accessible experience.

A toolkit approach also supports a personalized and accessible experience with the mindfulness technology. Personalization in technology is a meaningful way to support ownership of technology \cite{Cochrane21}. Designers should consider the user's feelings and preferences, not only for physical artifacts but also in the digital aspects, such as providing users with a choice of background color for an app.  These seemingly small choices can create a more personalized experience, provide users with a sense of control and ownership, and increase their engagement with mindfulness practice. As Jack Kornfield's work emphasizes the importance of finding a practice that suits the individual, recognizing that everyone's journey and preferences might differ. \cite{kornfield2009path}. Additionally, despite concerns regarding the potential drawbacks of AI, it can be employed as a helpful guide in this context. By integrating AI into mindfulness \cite{Kinkhabwala2020-jg, Kumar2023-ua}, technologies could offer personalized guidance and support, adapting to users' experience levels and individual needs. For example, an AI-powered chatbot can help beginners integrate mindfulness into daily activities by offering context-relevant examples tailored to their needs.

\subsubsection{Reappropriating non-mindfulness technology for mindfulness}

Our findings (\ref{findings - non-mindfulness tools}) suggested that mindfulness practitioners often utilized technologies that are not explicitly designed for mindfulness to facilitate their practice. We call designers to welcome such creative appropriations when the user acts as an "everyday designer" \cite{Ron2008}, integrating an ecosystem of technologies for their idiosyncratic practice. Reappropriating familiar tools one has already used, such as scheduling apps and timers, can be effective aids in promoting mindfulness. This also echoes the idea of "making strange" \cite{Loke2013-wz, loke2007making, loke2007making}, allowing users to reframe everyday tools for various activities, including mindfulness practices. 

Despite the availability of numerous wearables and mobile apps specifically designed to support mindfulness, designers should also explore existing technologies and determine how they can be repurposed to encourage mindfulness. For example, while productivity tools traditionally emphasize efficiency and strict scheduling, designing mindfulness into them (e.g., prompting a short mindfulness exercise, especially during prolonged periods of work) could foster well-being and reflection amidst daily tasks. In the case of a digital calendar, using it to schedule a mindfulness session may seem redundant with notifications, but without scheduling these sessions in one's agenda, consistent mindfulness may rely much on one's memory and self-discipline. Besides, by integrating the mindfulness concept, and the present awareness into technologies that users are already using, designers can promote more widespread adoption and diverse practice in different contexts. For example, non-mindfulness-specific platforms, such as music and video platforms, can be effectively utilized for mindfulness practice as they have already been widely used in one's routine.

\subsubsection{Minimizing the negative effects of technology on mindfulness}
Our findings (\ref{findings - tech nonuse}) suggest that designers should consider minimizing the negative effects of technology by promoting a healthy relationship between mindfulness and technology \cite{Levy2016-jh}. In contrast to digital well-being tools that focus on limiting screen time \cite{Hiniker2016-sh} or discouraging the use of certain applications \cite{Kim2019-rv}, our findings point (similar to \cite{Lukoff2018-kr}) toward promoting intentional and mindful technology use. This support can happen before, during, or after people interact with technology for mindfulness practice. During the practice of mindfulness, digital devices could automatically turn on silent mode to keep a tranquil environment for users to practice mindfulness based on the phone status (e.g., if the user is playing a guided mindfulness practice). To provide concrete suggestions for this design implications, we use the phone as an example because of its prevalence in people's daily life. 

Phones can provide lots of mindfulness resources, and offer reminders and other tools for mindfulness. However, some of our participants indicated a conflict around using their phone for mindfulness which adds to their overall phone use. A design that simplifies or shortens the steps of reaching the needed support (e.g, starting a meditation timer) for example, through one-click shortcuts \footnote{https://support.apple.com/guide/shortcuts/} could reduce potential distractions from one's phone. This design may be especially valuable for practicing mindfulness for better sleep, given that sleep can be facilitated by mobile mindfulness resources \cite{Huberty2021-dp} and at the same time be disrupted by excessive phone use time \cite{Li2015-bn}.   

Another suggestion coming from our participants (in \ref{findings - mindfulness as an object}) is transforming the use of technology into a mindful activity. Rather than seeing technology as a separate object outside the mindfulness practice, researchers may develop mindfulness resources to teach people to use the mindfulness principle to become aware of their technology use. This echos what Levy \cite{Levy2016-jh} teaches in his book \textit{Mindful Tech} and the Mindphone intervention that prompts mental check-ins or writing reflections whenever users unlock their phone \cite{Terzimehic2022-xh}. 

A third way to minimize technology's negative effects is to integrate mindfulness exercises as digital well-being interventions. Similar to applications that encourage taking micro-breaks from screens for physical health to reduce eye strain (e.g., \textit{EyeLeo}) and to stretch (e.g., \textit{Stretchly}), a mindfulness break can recommend taking an outdoor walk or doing a quick breathing exercise to create a pause and leave space for mindful decisions. One such example is \textit{Microsoft Viva Insights}\footnote{https://www.microsoft.com/en-us/microsoft-viva/insights}, a workplace analytics tool that incorporates reflection features and interfaces with \textit{Headspace} for guided meditations, aiming to increase productivity by promoting mental well-being. Another example is Zenscape, a tangible system that offers mini zen garden activities, engaging users in mindful micro-breaks from screen work \cite{Khot2022-bm}. We suggest that designers incorporate mindfulness exercises as an internal, soft, and non-judgemental digital well-being strategy to potential technology overuse, which extends previous work of enhancing user agency in technology use with internal and abstract support \cite{Zhang2022-ow} and may also reduce the temptation to re-engage with the devices \cite{aranda2018toward}.

\subsubsection{Bringing community support into personal mindfulness practice}
\label{implication: bringing community support}
Mindfulness is mostly seen as an individual and personal journey: it is about one's personal mind, connected with their body, life, and activities. But this personal journey doesn't happen in a social vacuum. Our participants (in \ref{findings - use technology for community}) talked about having meaningful conversations with others in group meditation sessions over Zoom and in-person, a community that they connect with, and teachers they look up to. Together, these experiences make this personal journey bigger and more meaningful. Future studies may explore ways to support consistent mindfulness practice by bringing community support into personal mindfulness practice. One example of that is to understand and explore more the use of computer-mediated communication (CMC) technologies in the group mindfulness practice \cite{derthick2014understanding} and enhance the support. 

A community-based approach to designing for mindfulness opens up more opportunities for supporting mindfulness through group effort, by learning from, being inspired by, and getting support from fellow practitioners. Future research could explore ways of strengthening this connection. For example, Mindful Garden used augmented reality and sensors to create shared mindfulness experiences among two meditation partners to enhance feelings of connectedness and relatedness \cite{Liu2022-wz}. Furthermore, although individuals may learn and practice mindfulness on their own, connecting to a teacher allows people to get more personalized guidance and help along the journey. The importance of findings a teacher to support mindfulness practice has been both emphasized by our participants and in previous research \cite{Van_Aalderen2014-rw}. Future research may explore how to help individuals find and connect to teachers in various forms (e.g., in-person, live stream \cite{Guo2022-zm, Li_Guo_LiveStream}, resources) and the benefits and downsides of each form for everyday mindfulness practice. Beyond supporting personal mindfulness, future work could also look at the ways in which technologies can help build and enhance \textit{interpersonal} mindfulness. For example, a design intervention that slows down the pace of polarizing online conversations can help conversation partners reflect on and be more thoughtful about messages they write \cite{Masrani2023-si}. 

\subsection{Limitations and Future Work}
This paper presents in-depth accounts of everyday mindfulness practice from 20 practitioners, all having long-term experience in mindfulness (4 to 20+ years). Our findings are therefore limited to the experiences and practices of these advanced mindfulness practitioners who are beyond the stage of learning mindfulness. Many of them being teachers, they provided useful suggestions for finding the right forms of mindfulness practice based on their interactions with their novice students. However, our findings may not generalize to the perspectives of the novices. Furthermore, we did not investigate differences between practitioners based on their levels or years of experience, e.g., less/more than 10 years versus over 10 years of practice. To address this, future research could explore these aspects more comprehensively and develop personalized support strategies for practitioners based on their level of experience. Moving forward, one future direction is to conduct longitudinal studies with novice practitioners and track how they develop their skills and maintain the practice with technology use over the long term \cite{Odom2019-yt}. Another direction is to design and evaluate tools that support mindfulness in everyday activities that incorporate personal goals, experiences, and daily schedules (e.g., in videoconference meetings \cite{Wu-auto2024} and casual walks \cite{Tan2023-mindfulmoment}). 

\section{Conclusion}
In this paper, we present findings from a semi-structured interview study with 20 experienced mindfulness practitioners, with the aim to gain in-depth insights into individualized everyday mindfulness practice, motivations, experienced benefits, and challenges of keeping consistent mindfulness practice, and the role of technology in the process. This work contributes to an in-depth understanding of everyday mindfulness practices beyond meditation, technologies used in the practice, and guiding the design of mindfulness technologies that are better situated in practitioners' daily lives for consistent mindfulness.

%%
%% The acknowledgments section is defined using the "acks" environment
%% (and NOT an unnumbered section). This ensures the proper
%% identification of the section in the article metadata, and the
%% consistent spelling of the heading.
\begin{acks}
We would like to thank all mindfulness practitioners for being willing to share their experiences with us. We would also like to thank Lee Humphreys and Katherine Sender for their valuable feedback on the initial draft of this paper.
\end{acks}

%%
%% The next two lines define the bibliography style to be used, and
%% the bibliography file.
\bibliographystyle{ACM-Reference-Format}
\bibliography{sample_base}

%%
%% If your work has an appendix, this is the place to put it.
\appendix

\section{Categories, descriptions, and examples of codes}
\label{table-codes}
\begin{table}[h]
\centering
\begin{tabular}{m{4cm}m{6cm}m{6cm}}
\hline
\textbf{Category} & \textbf{Description} & \textbf{Examples of Specific Codes} \\ 
\hline
Definitions of Mindfulness & Describes how participants define mindfulness in both formal and informal contexts. & Paying attention to the present moment; doing one thing at a time; destressing and relaxation \\ 
\hline
Formal and Informal Mindfulness Practice & Describes both formal and informal mindfulness practices & Sitting meditation; Mindful eating; Mindful walking; yoga; attention to breath; mindfulness phone use; mindful driving \\ 
\hline
Benefits of Mindfulness & Short-term and long-term benefits experienced by participants & Better sleep; reducing chronic pain; reduced stress; better mental clarity; increasing emotional stability; improved interpersonal relationships \\ 
\hline
Barriers to Mindfulness & Challenges faced by participants in maintaining consistent mindfulness practice. & Busy schedules; difficulty finding time and space; strong emotions or physical pain; habitual patterns; \\ 
\hline
Technology Use & The role of digital tools in supporting mindfulness practice & Use of mindfulness apps like Headspace for guided sessions; attending online group sessions via Zoom; reminders; use technology as a focus object for mindfulness practice \\ 
\hline
Technology Non-use & Describes scenarios where participants choose to limit or avoid technology to enhance mindfulness or avoid distractions. & Silencing notifications during practice; applying a black and white mode; limiting time spent on devices; removing devices \\ 
\hline
Strategies for Consistent Practice & Strategies developed by participants to overcome barriers and keep consistent mindfulness practice. & Practice short mindfulness exercises in morning/night routine; joining mindfulness online/offline community; seeking help from a teacher \\ 
\hline
\end{tabular}
\caption{Overview of categories, descriptions and example of codes in the data analysis }
\end{table}

\end{document}